\documentclass[a4paper,fleqn,usenatbib,useAMS]{mnras}

\usepackage{graphicx}	
\usepackage{amsmath}	
\usepackage{amssymb}	
\usepackage{multicol}        
\usepackage{bm}		
\usepackage{pdflscape}	
\usepackage[encapsulated]{CJK}
\usepackage{ucs}
\usepackage[utf8x]{inputenc}

\usepackage[T1]{fontenc}
\usepackage{ae,aecompl}

\usepackage{newtxtext,newtxmath}

\title[{\it Chandra} observations of IC\,4593]{{\it Chandra} observations of the planetary nebula IC\,4593}
\author[Toal\'{a} et al.]{
J.A.\ Toal\'{a}$^{1}$\thanks{E-mail:\,j.toala@irya.unam.mx}, 
M.A.\ Guerrero$^{2}$, 
L.\ Bianchi$^{3}$, 
Y.-H.\ Chu$^{4}$ and 
O.\,De Marco$^{5}$\\
  $^1$Instituto de Radioastronom\'{i}a y Astrof\'{i}sica (IRyA), UNAM Campus Morelia, Apartado postal 3-72, 58090 Morelia, Michoacan, Mexico \\
  $^2$Instituto de Astrof\'{i}sica de Andaluc\'{i}a (IAA-CSIC), Glorieta de la Astronom\'{i}a S/N, 18008 Granada, Spain \\
  $^3$Department of Physics and Astronomy, The Johns Hopkins University, Baltimore, MD, USA \\
  $^4$Institute of Astronomy and Astrophysics, Academia Sinica (ASIAA), 
      Taipei 10617, Taiwan, Republic of China \\
  $^5$Astronomy, Astrophysics and Astrophotonics Research Centre, Macquarie University, Sydney, NSW 2109, Australia
}

\pubyear{2019}

\begin{document}
\label{firstpage}
\pagerange{\pageref{firstpage}--\pageref{lastpage}}
\maketitle

\begin{abstract}

The ACIS-S camera on board the \emph{Chandra X-ray Observatory} has
been used to discover a hot bubble in the planetary nebula (PN)
IC\,4593, the most distant PN detected by \emph{Chandra} so far.  The
data are used to study the distribution of the X-ray-emitting gas in
IC\,4593 and to estimate its physical properties.  The hot bubble has
a radius of $\sim$2$^{\prime\prime}$ and is found to be confined
inside the optically-bright innermost cavity of IC\,4593.  The X-ray
emission is mostly consistent with that of an optically-thin plasma
with temperature $kT\approx0.15$~keV (or
$T_\mathrm{X}\approx1.7\times10^{6}$ K), electron density
$n_\mathrm{e}\approx15$~cm$^{-3}$, and intrinsic X-ray luminosity in
the 0.3--1.5~keV energy range $L_\mathrm{X}=3.4\times10^{30}$
erg~s$^{-1}$.  A careful analysis of the distribution of hard
($E>$0.8~keV) photons in IC\,4593 suggests the presence of X-ray
emission from a point source likely associated with its central star
(CSPN).  If this were the case, its estimated X-ray luminosity would
be $L_\mathrm{X,CSPN}=7\times10^{29}$~erg~s$^{-1}$, fulfilling the
log$(L_\mathrm{X,CSPN}/L_\mathrm{bol})\approx-7$ relation for
self-shocking winds in hot stars.  The X-ray detection of the CSPN
helps explain the presence of high-ionisation species detected in the
UV spectra as predicted by stellar atmosphere models.
  
\end{abstract}





\section{INTRODUCTION}
\label{sec:intro}

The unprecedented angular resolution of the \emph{Chandra X-ray
  Observatory} has revealed in exquisite detail the hot interiors of
planetary nebulae (PNe).  The first \emph{Chandra} observations of PNe
already revealed that hot bubbles are contained within their inner
rims \citep[see the cases of BD$+$30$^{\circ}$3639 and
  NGC\,6543;][]{Kastner2000,Chu2001}.  These hot bubbles form as the
result of interactions of the fast wind from the central star (CSPN)
with material ejected during the previous asymptotic giant branch
(AGB) phase.  This wind-wind interaction creates an
adiabatically-shocked region or hot bubble with temperatures that are
determined by the CSPN terminal wind velocity as $T\sim v_{\infty}^2$
\citep[][]{Dyson1997}. Typical wind velocities from CSPN are reported
to be $v_{\infty}\gtrsim500-4000$~km~s$^{-1}$
\citep[e.g.,][]{Guerrero2013}, producing shocked plasma at
X-ray-emitting temperatures in excess of 10$^6$~K.  These general
X-ray properties of hot bubbles in PNe have been largely confirmed by
the \emph{Chandra} Planetary Nebula Survey
\citep[ChanPlaNS;][]{Kastner2012,Freeman2014}, a volume-limited survey
of a sample of nearby ($d<1.5$~kpc) PNe. Extended X-ray emission has a
detection rate among PNe $\sim$30\%, with particularly high prevalence
among compact ($r<0.2$~pc) PNe with closed-shell morphology.

\emph{Chandra} has also unveiled unexpected point-sources of hard
X-ray emission ($>$0.5 keV) from CSPNe \citep{Guerrero2001}, whose
presence in a significant number of them has been confirmed by
ChanPlaNS \citep{Montez2015}. Whereas some of these hard X-ray sources
can be attributed to the coronal emission from a dwarf or giant
late-type companion \citep{Montez2010} or even to accretion onto a
compact companion \citep{Guerrero2019}, shocks in fast winds as in OB
and Wolf-Rayet stars can also be the cause of this hard X-ray emission
\citep{Guerrero2001}, especially among CSPNe with powerful fast
stellar winds. As these stellar winds are radiatively driven and thus
inherently unstable, small-scale inhomogeneities can lead to strong
stochastic shocks within the dense wind layers and heat material to
X-ray-emitting temperatures \citep{LW1980,GO1995}. Single OB stars are
typically detected in X-rays ($L_{\rm X}$=$10^{31}$--10$^{33}$
ergs~s$^{-1}$, $kT$=0.2--0.5 keV) with X-ray to bolometric luminosity
ratios $L_{\rm X}/L_{\rm bol}\sim10^{-7}$ \citep{B1997,S2006,NGO2018}.

\begin{figure*}
\begin{center}
  \includegraphics[angle=0,width=0.95\linewidth]{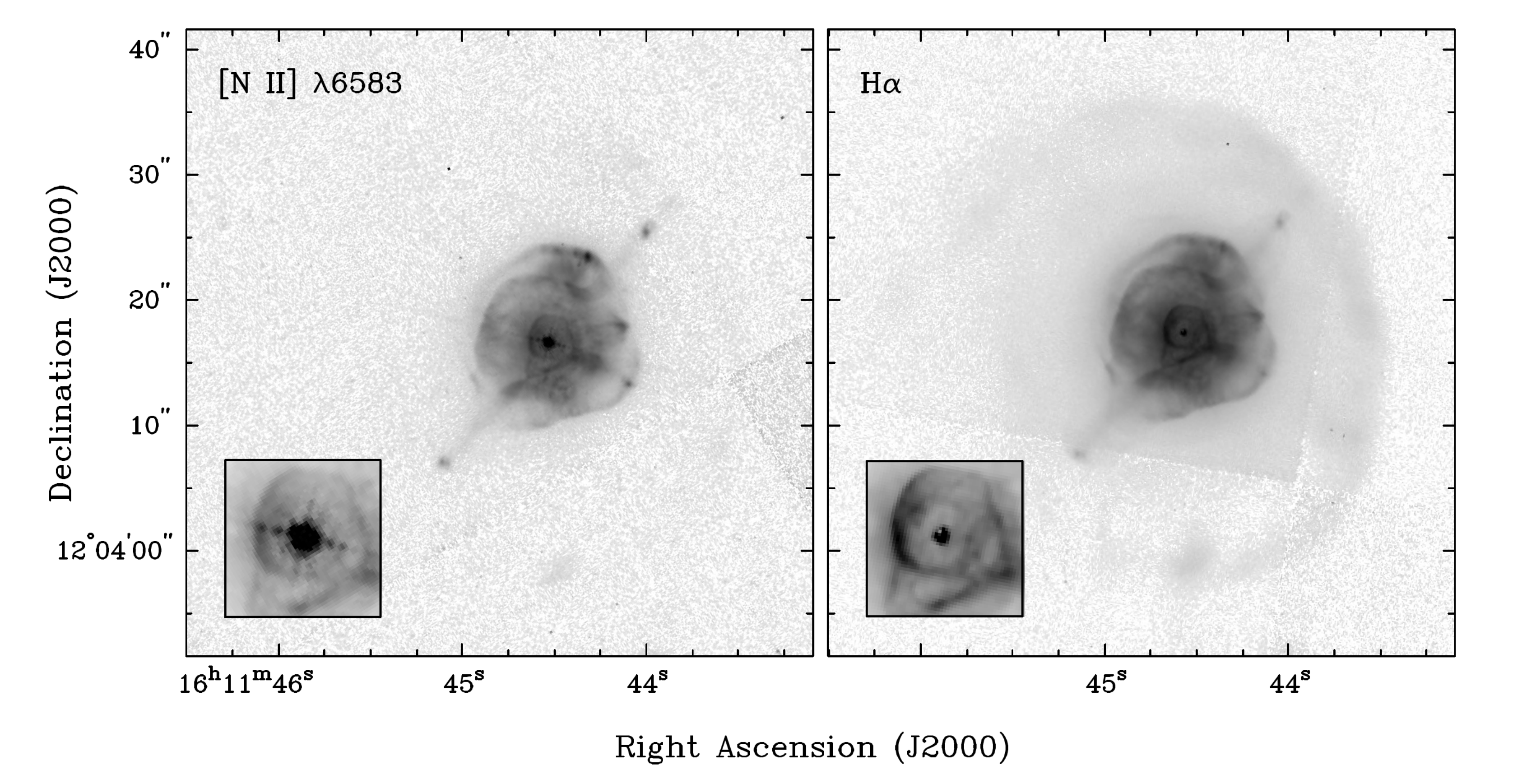}
\label{fig:opt}
\caption{ \emph{HST} WFPC2 [N~{\sc ii}] $\lambda$6583 ({\it left}) and
  H$\alpha$ ({\it right}) images of the main nebular region of
  IC\,4593. The inset shows the innermost
  5$^{\prime\prime}\times$5$^{\prime\prime}$ region.}
\end{center}
\end{figure*}

Further evidence for wind shocks in OB stars is provided by P-Cygni
profiles of the so-called super-ions, species with ionisation
potentials (IP) well above that of He~{\sc ii} (IP=54.4 eV).
Super-ions in OB stars cannot be abundantly produced by
photo-ionisation because the stellar continuum shortwards of 228
\AA\ is largely absorbed by He~{\sc i} for bound-free transitions.
Thus, Auger ionisations by X-ray photons generated by wind shocks
provide the only plausible mechanism to produce super-ions
\citep{CO1979}.  Detailed modeling of UV and optical spectra of H-rich
CSPNe indeed indicates that the inclusion of X-rays in the model
atmosphere calculation is absolutely necessary to reproduce with a
uniquely consistent effective temperature the S~{\sc vi} (IP of
S\,{\sc v}~=~72.6~eV), N~{\sc v} (IP of N\,{\sc iv}~=~77.5~eV), and
O~{\sc vi} (IP of O\,{\sc v}~=~113.9~eV) doublets for stars with
$T_\mathrm{eff}\leq$45,000~K, while it improves the spectral fit for stars
with $T_\mathrm{eff}$=55,000--80,000~K \citep{Herald2011}.  Among the stars
with these spectral properties, the CSPN of IC\,4593 stands out for
its noticeable O~{\sc vi} P Cygni profiles ($v_\infty = 1,150$
km~s$^{-1}$) and low effective temperature, $\simeq$41,000~K.  It also
stands out for its wind variability, detected both in \emph{IUE}
\citep{PP1995} and \emph{FUSE} \citep{Guerrero2013} spectra, with
ripples moving bluewards in the troughs of P Cygni profiles in various
UV lines with timescales of a few hours similar to those of its
pervasive optical lines \citep{dM2004} and photometric variations
\citep{Miszalski2009}.


IC\,4593 exhibits a complex optical multiple-shell morphology with a
roughly elliptical inner rim with average radius
$\sim2^{\prime\prime}$, a roundish bright filamentary outer shell
$\sim16''$ in size, two prominent jet-like features protruding from
the inner shell towards NW and SE, and an asymmetrical arc-like outer
shell that extends $\sim20''$ from the CSPN.  In addition to this main
jet-like feature, there are others aligned along different directions,
which seem to be stagnated at the edge of the shell, as implied by
their negligible systemic velocities \citep{Corradi1997}. All these
features can be appreciated in Figure~\ref{fig:opt}.  Furtheremore,
there is an additional halo
$\sim$130$^{\prime\prime}$$\times$120$^{\prime\prime}$ in size
\citep[see figure~2 in][]{Corradi1997}, whose fragmented and
asymmetrical morphology is indicative of interaction with the
interstellar medium
\citep[see][]{Zucker1993,Villaver2003,Villaver2012}.

The complex morphology of the main nebula of IC\,4593 is suggestive of
the action of different shaping agents. The current fast stellar wind
would be responsible of the formation of the inner rim, whereas the
collimated outflows protruding along different directions
\citep{Corradi1997,OConnor1999} would shape the outer shell.
Actually, \citet{Corradi1997} suggest that the inner rim might have
been pierced by the jets ending with the pair of knots aligned along
the SE-NW direction, which may be associated with bipolar jets,
although this would require an inclination angle of the outflow with
respect to the plane of the sky $\lesssim$15$^{\circ}$ because no
kinematical signature is detected in their high-dispersion spectra.
Episodic rotating jets would be a natural consequence of a precessing
accretion disk around a binary companion, as expected after the
evolution of a binary system through a common envelope phase
\citep[see][and references therein]{Zou2019}.
Indeed, the presence of a binary system at the core of IC\,4593 has 
been suggested through radial velocity measurements of absorption 
lines of He\,{\sc ii} \citep[e.g.,][]{dM2004} making this system a 
candidate for binary interactions and for the emission of X-rays 
from the coronal activity of a dwarf late-type companion.

Here we present Cycle~13 \emph{Chandra} ACIS-S observations of
IC\,4593, which was not originally included in the ChanPlaNS sample
because of its large distance, $d$=2.4$^{+0.6}_{-0.4}$~kpc
\citep{BJ2018}. We use our {\it Chandra} observations to unveil the
presence of extended X-ray emission within IC\,4593 and to search for
possible X-ray emission from its central star or a binary companion. 
The paper is organized as follows: 
the observations and data analysis are presented in Section~2, 
the results discussed in Section~3, and a summary presented in 
Section~4.

\section{Observations and data analysis}

\begin{figure*}
\begin{center}
\includegraphics[angle=0,width=\linewidth]{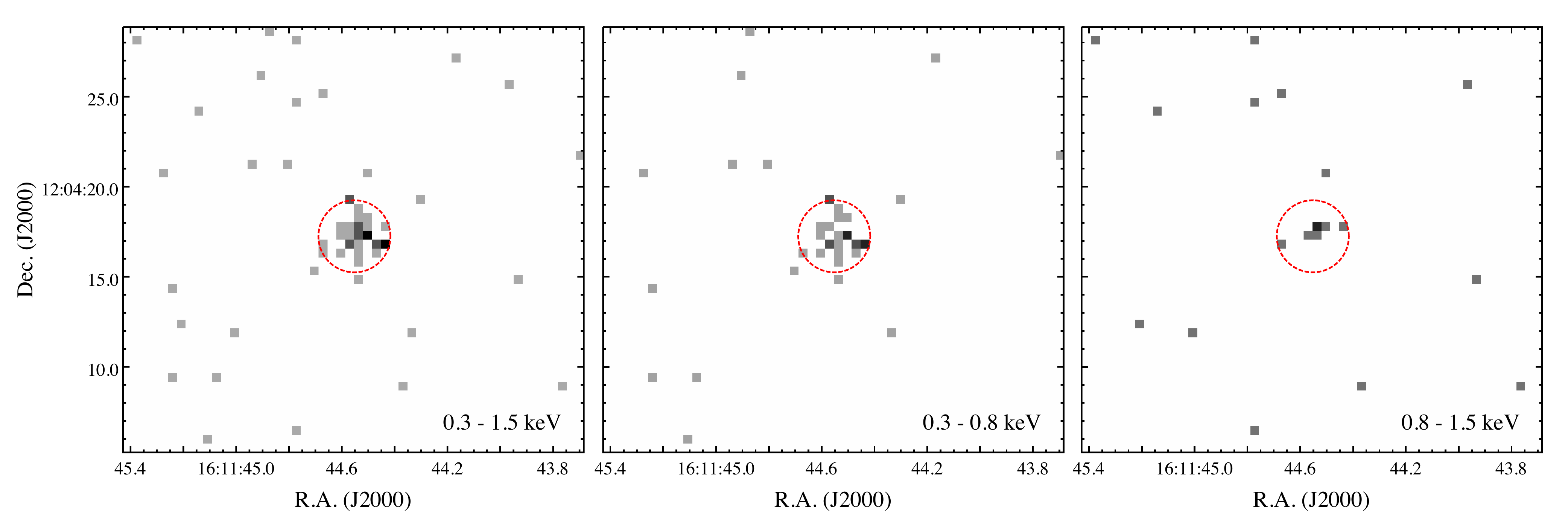}
\label{fig:figx}
\caption{ \emph{Chandra} ACIS-S event images of IC\,4593 in the
  0.3--1.5~keV (left), 0.3--0.8~keV (center), and 0.8--1.5 (right)
  energy bands.  The red dashed circle is centered on the CSPN of
  IC\,4593 and has an angular radius of 2$^{\prime\prime}$.  The
  images have been produced using the natural ACIS-S pixel size of
  0\farcs5.  }
\end{center}
\end{figure*}


IC\,4593 was observed by \emph{Chandra} on 2012 January 2 during Cycle
13 (PI: M.A.\ Guerrero; Obs.\ ID.\ 13654). The observations were
performed with the back-illuminated CCD\,3 of the ACIS-S array in the
VFAINT mode for a total exposure time of 80~ks.  After processing the
data with the \emph{Chandra} Interactive Analysis of Observations
\citep[{\sc ciao}; version 4.9;][]{Fruscione2006}, the resulted total
useful time is 76.3~ks. There were no periods of time in which the
data were affected by high-level backgrounds.  The \emph{Chandra}
observations detect X-ray emission at the location of IC\,4593, with a
total ACIS-S3 count number of 31$\pm$6~counts in the energy range
0.3-1.5 keV, which corresponds to a net ACIS-S3 count rate of
0.38$\pm$0.07 counts~ks$^{-1}$ in this same energy range. We
investigate next the spatial distribution and spectral properties of
this X-ray emission.

\subsection{Spatial properties of the X-ray emission}

\begin{figure*}
\begin{center}
  \includegraphics[angle=0,width=\linewidth]{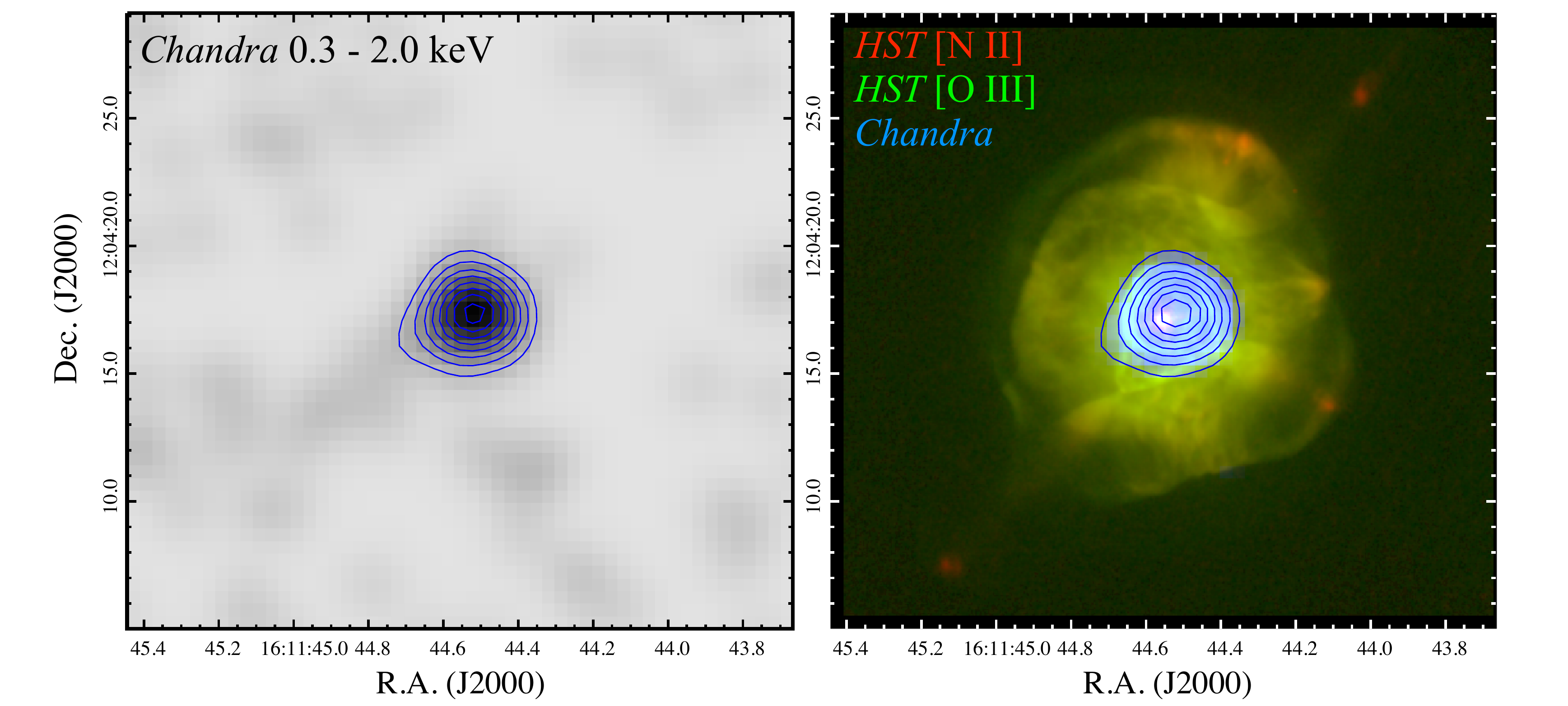}
\label{fig:fig}
\caption{ Left: Smoothed {\it Chandra} ACIS-S image of IC\,4593
  created for the 0.3--1.5~keV energy range.  Right:
  color-composite optical and X-ray image. The contours correspond to
  the X-ray emission.}
\end{center}
\end{figure*}

To examine the spatial distribution of the X-ray emission, we created
an image of the X-ray events in the 0.3--1.5~keV energy range (left
panel of Figure~\ref{fig:figx}), which is the energy range in which
most of hot bubbles in PNe have been detected \citep[see][and
  references therein]{Ruiz2013,Kastner2012}.  The image, produced
using the natural ACIS-S pixel size of 0\farcs5, reveals the presence
of extended emission spatially confined within the inner rim of
IC\,4593, which is represented by a circular aperture with an angular
radius of 2$^{\prime\prime}$.

To further confirm the spatial correspondence between the diffuse
X-ray emission of IC\,4593 and its optical inner rim, we created a
smoothed image of the X-ray emission from IC\,4593 in the 0.3--2.0~keV
band using the {\sc ciao} task {\it aconvolve}.  The image is smoothed using
a circular Gaussian kernel with size up to 4 pixels
($\approx2^{\prime\prime}$) and a fast-Fourier transform (FFT)
convolution method. The final smoothed image is presented in the left
panel of Figure~3 and, for comparison with the optical morphology, a
color-composite image of IC\,4593 using the \emph{HST} [N\,{\sc ii}]
and [O\,{\sc iii}] narrowband images and the smoothed X-ray image is
presented in the right panel. Figure~3 right panel shows that the
X-ray emission fills the innermost cavity of IC\,4593.

\subsection{Spectral properties of the X-ray emission}

The spectrum of the X-ray emission from IC\,4593 was extracted using
the {\sc ciao} task {\it specextract} from a region with a radius of
2$^{\prime\prime}$ enclosing the inner rim.  A background spectrum was
extracted from a region with no contribution from background X-ray
sources.  The resultant background-subtracted ACIS-S spectrum of
IC\,4593 is shown in Figure~4. Given the small number of counts, the
spectrum quality is notably poor. Nonetheless, its spectral shape is
consistent with the emission from a low-temperature optically-thin
plasma typical of PN hot bubbles \citep{Montez2005,Ruiz2013}; the
spectrum is soft and peaks between 0.5--0.6 keV with apparent line
emision features around $\sim$0.4 keV and $\lesssim$1.0 keV.  The
dominant peak can be attributed to the O~{\sc vii} triplet, while the
emission around 0.4~keV might be due to the contribution from C~{\sc
  iv} and/or N~{\sc vi} lines.  No emission is detected beyond
2.0~keV.

\begin{figure}
\begin{center}
  \includegraphics[angle=0,width=0.9\linewidth]{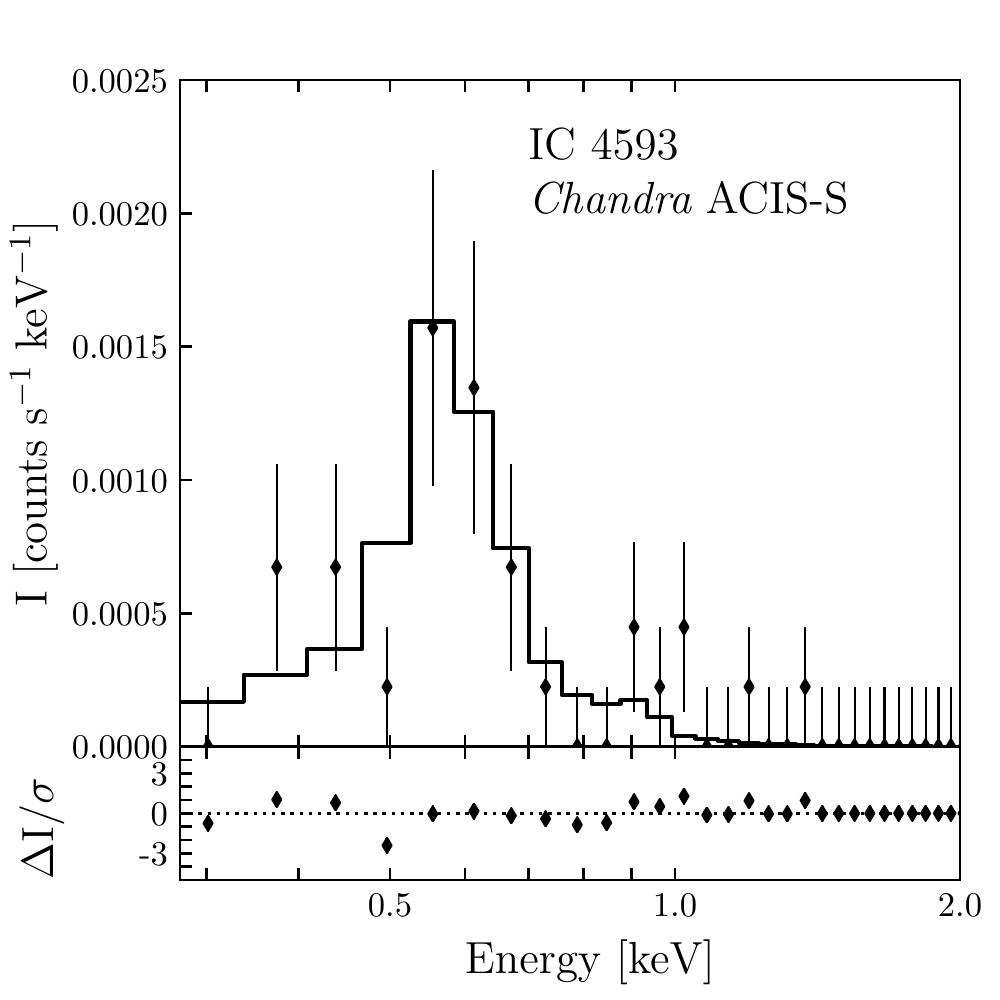}
\label{fig:spec}
\caption{
Background-subtracted \emph{Chandra} ACIS-S spectrum of the extended X-ray 
emission in IC\,4593. 
The error bars correspond to 1$\sigma$ uncertainty.  
The solid line plots a model that fits the observed spectrum reasonably well. 
The bottom panel shows the residuals of the model.
}
\end{center}
\end{figure}

Since the data quality does not warrant a standard spectral fit using
the reduced $\chi^2$ method, we have rather qualitatively compared the
observed background-subtracted spectrum with a number of synthetic
{\it apec}\footnote{A collisionally-ionised diffuse gas model
  calculated from the AtomDB atomic database
  (\url{http://atomdb.org/}) which is included in XSPEC.}
optically-thin plasma X-ray spectra modelled using XSPEC version
12.10.1 \citep[][]{Arnaud1996}.  A fixed hydrogen column density
$N_\mathrm{H}=4.3\times10^{20}$~cm$^{-2}$, as estimated by the NASA
HEASARC $N_\mathrm{H}$
tool\footnote{\url{https://heasarc.gsfc.nasa.gov/cgi-bin/Tools/w3nh/w3nh.pl}},
was adopted as the absorption column. We note that this value is
consistent with the $E(B-V)$ value of 0.06~mag reported by
\citet{Herald2011}.  A model with plasma temperature
$T_\mathrm{X}\lesssim0.15$~keV (plotted in solid line in Figure~4)
describes reasonably well the ACIS-S3 spectrum.  Higher plasma
temperatures ($T_\mathrm{X}\simeq0.20$ keV) displace the dominant
spectral feature at $\sim$0.55~keV towards higher energies, whereas
lower plasma temperatures ($T_\mathrm{X}\simeq0.10$ keV) produce
larger emission at lower energies.  The estimated observed flux of the
model is $f_\mathrm{X} \approx$
3.3$\times$10$^{-15}$~erg~s$^{-1}$~cm$^{-2}$, which corresponds to an
unabsorbed X-ray flux of $F_\mathrm{X} \approx$
5.0$\times$10$^{-15}$~erg~s$^{-1}$~cm$^{-2}$ and a luminosity of
$L_\mathrm{X} \approx 3.4\times$10$^{30}$~erg~s$^{-1}$.  For the X-ray
temperature range between 0.1 and 0.2 keV described above, models with
X-ray luminosities in excess of $6\times$10$^{30}$~erg~s$^{-1}$ and
below $2\times$10$^{30}$~erg~s$^{-1}$ result in predicted spectral
shapes that deviate significantly from the observed spectrum. The
normalization parameter $A$, estimated to be
3.9$\times$10$^{-6}$~cm$^{-5}$, was used to derive an electron density
of $n_\mathrm{e}\sim$15~cm$^{-3}$ for the hot bubble in
IC\,4593\footnote{The normalization parameter can be expressed as
  $A=10^{-14}\int n_\mathrm{e}^{2}dV/4 \pi d^{2}$, where $d$ is the
  distance and $V$ is the volume of the X-ray-emitting region.}.

\subsection{X-ray emission from the CSPN?}

The comparison between the X-ray spectrum of IC\,4593 and the
\emph{apec} thin-plasma emission model overplotted in Figure~4 reveals
some emission excess at $\sim$0.4 keV and $\sim$0.9 keV.  The spatial
distribution of the X-ray emission below 0.5 keV is consistent with
that of the bulk of X-ray emission, but that of the emission above 0.8
keV shows a remarkable clustering at the position of the CSPN of
IC\,4593, with 5 photons within a radius of 0\farcs8 (Figure~2-{\it
  left}).  We have used the {\sc ciao} task {\it simulate$\_$psf}, in
conjunction with {\sc marx} 5.5.0 \citep{Davis_etal2012}, to simulate
the \emph{Chandra} point spread function (PSF) for a monochromatic
spectrum at 1.0 keV at the location of the CSPN of IC\,4593.  This PSF
has been used to run a number of 100,000 Monte Carlo simulations of a
detection of 7 counts.  It is found that a percentage of these
simulations above 95\% result in a spatial distribution of the counts
that conforms to that shown in the rightmost panel of Figure~2.

The distribution of the X-ray emission with energies above 0.8 keV
from IC\,4593 thus presents tantalizing evidence for emission from its
CSPN.  A strong temperature gradient of the X-ray-emitting plasma in
the hot bubble might be claimed to produce a similar spatial
distribution, but such temperature gradients have not been observed so
far in any PN \citep{Yu_etal2009}.  Interestingly, the X-ray spectrum
of other CSPNe whose hard X-ray emission has been associated with a
shock in origin, such as NGC\,6543 and NGC\,6826, show similar
spectral features \citep{Guerrero2001,Ruiz2013}.

This feature may be indicative of the presence of a plasma with an
over-abundance of Ne, or with tempertures $\sim$10$^7$ K \citep[as for
  the CSPN of NGC\,6543,][]{Guerrero2001}. Assuming a plasma at this
temperature, the count rate above 0.8 keV of 0.09$\pm$0.03
counts~ks$^{-1}$, implies an unabsorbed X-ray luminosity in the range
(2--16)$\times$10$^{29}$ erg~s$^{-1}$.

\section{Discussion}

IC\,4593 is the farthest PN detected in X-rays by the \emph{Chandra
  X-ray Observatory}\footnote{
The \emph{Chandra} dramatic decrease of effective area at low energies due to 
the buildup of a contaminant onto the CCD window may sadly render this record
definitive \citep{Guerrero2020}, although we note that there is currently a 
search of X-ray-emitting PNe in globular clusters using archival \emph{Chandra}
data that might result in interesting findings \citep[][]{Montez2020}.}.
The detection of X-ray emission from IC\,4593 by \emph{Chandra} presented 
here can be argued to be marginal, but we note that the total detected 
photons (31$\pm$6~counts) translates into a $\sigma \approx 5$ detection.  
Moreover, our results are
consistent with other X-ray detections of hot bubbles in PNe: the
X-ray emission of IC\,4593 is soft with its bulk in the 0.3--1.0~keV
energy range, the emission is extended and is confined within the
inner rim of the nebula, and the spectral properties of the
X-ray-emitting plasma lie within the observed values of other hot
bubbles ($T_\mathrm{X}\approx1.7\times10^{6}$~K,
$n_\mathrm{e}\approx15$~cm$^{-3}$, and
$L_\mathrm{X}\approx10^{30}$~erg~s$^{-1}$).  Furthermore, the presence
of a narrow absorption component on the O~{\sc vi} $\lambda$1032
\AA\ P Cygni profile of the \emph{FUSE} spectrum reported by
\citet{Guerrero2013} is suggestive of a conductive or mixing layer
between the nebular material and the adiabatically-shocked hot bubble
\citep[][]{Iping2002,Gruendl2004,Ruiz2013}, which is the main physical
mechanism suggested to produce the soft X-ray emission in PNe
\citep[][]{Steffen2008,Fang2016,Toala2018}.

At a distance of 2.4~kpc, the inner rim of IC\,4593 has a physical
radius of 0.023~pc, i.e., it harbors one of the most compact hot
bubble detected in X-rays together with
BD$+$30$^{\circ}$3639, IC\,418, and NGC\,7027
\citep[][]{Kastner2012,Freeman2014}. 

The spectral properties of the X-ray emission from IC\,4593 and
especially the spatial distribution of the X-ray emission harder than
0.8 keV are suggestive of a point-source at its CSPN.  The X-ray
luminosity of this emission implies an X-ray to stellar bolometric
luminosity\footnote{The $L_\mathrm{bol}$ value derived for the CSPN of
  IC\,4593 from atmosphere model fits by \citet{Herald2011} has been
  rescaled to the distance used here.}  ratio
log$(L_\mathrm{X,CSPN}/L_\mathrm{bol})$ in the range between --7.0 and
--7.9. This ratio is consistent with the usual value found for CSPNe
dominated by shocks in winds \citep[e.g.,][and references
  therein]{Montez2015}, in line with the presence of super-ions in the
stellar wind of IC\,4593 \citep{Herald2011,Guerrero2013}.  On the
other hand, the X-ray luminosity of the CSPN of IC\,4593 is at the
upper range for the coronal activity of dwarf late-type stars
\citep[see Figure~3 of][]{Guerrero2019}, which would be achieved by an
early K or late G dwarf companion with saturated activity
\citep[log$(L_\mathrm{X}/L_\mathrm{bol})=-3.56$;][]{Fleming1995}.  At
the distance of IC\,4593, our Sun would have $m_V=16.73$ mag and
$m_K=15.27$ mag, much weaker than the CSPN of IC\,4593, making its
optical or near-IR detection unfeasible. Therefore, the current
\emph{Chandra} observations cannot be used to assess the origin of
this X-ray emission, which cannot be attributed neither to a
magnetically active binary companion nor to shocks within its stellar
wind.  We note, however, that the shocks in the stellar wind of
IC\,4593 revealed by UV observations would produce X-ray emission at
the level and with the spectral properties suggested by the
\emph{Chandra} observations presented here.

\section{Summary}

We present the analysis of \emph{Chandra} ACIS-S3 observations of 
the PN IC\,4593.  
The observations detect diffuse X-ray emission confined within the 
inner rim of IC\,4593, which would correspond to a hot bubble powered 
by the wind from its CPSN.  
Although the number of counts is small, the spectral properties of 
this X-ray emission are consistent with those of other hot bubbles 
in PNe, with a temperature $T_\mathrm{X}\approx$1.7$\times$10$^{6}$~K 
and an electron density $n_\mathrm{e}$=15~cm$^{-3}$.  
The hot bubble of IC\,4593 is one of the smallest among PNe, 
besides BD$+$30$^{\circ}$3639, IC\,418, and NGC\,7027.

A careful inspection of the hard photons ($E>0.8$~keV) suggests the
presence of a central source very likely associated with the CSPN.
The X-ray luminosity of such a point-source implies a
log$(L_\mathrm{X,CSPN}/L_\mathrm{bol})$ of $-$7.4, very similar to
what is observed in CSPNe and OB and Wolf-Rayet stars with
self-shocking winds. 
Alternatively, the X-ray emission can be attributed to an early K or late G 
dwarf companion, but its presence cannot be confirmed through optical or 
near-IR observations.  
Deeper X-ray observations are needed to confirm the presence of 
an X-ray-emitting CSPN in IC\,4593 and to characterise its nature.  
Such measurements could easily be provided by future 
X-ray missions as \emph{Lynx} and \emph{Athena}.

\section*{Acknowledgements}
JAT and MAG are supported by the UNAM DGAPA PAPIIT projects IA100318
and IA100720. MAG acknowledges support from grant
PGC2018-102184-B-I00, co-funded with FEDER funds.
LB and YHC acknowledge support from NASA grant \emph{Chandra} GO2-13024B. 
YHC acknowledges the research grant
108-2112-M-001-045 from the Ministry of Science and Technology (MOST)
of Taiwan. 
This work has made extensive use of the NASA's Astrophysics Data System.
Based on observations made with the NASA/ESA Hubble Space Telescope, 
and obtained from the Hubble Legacy Archive, which is a collaboration 
between the Space Telescope Science Institute (STScI/NASA), the Space 
Telescope European Coordinating Facility (ST-ECF/ESA) and the Canadian 
Astronomy Data Centre (CADC/NRC/CSA).




\begin{thebibliography}{99}

  \bibitem[Arnaud(1996)]{Arnaud1996} Arnaud, K.~A.\ 1996, Astronomical
  Data Analysis Software and Systems V, 17

\bibitem[Bailer-Jones et al.(2018)]{BJ2018} Bailer-Jones, C.~A.~L.,
  Rybizki, J., Fouesneau, M., et al.\ 2018, VizieR Online Data
  Catalog, I/347
  
\bibitem[Berghoefer et al.(1997)]{B1997} 
Berghoefer, T.~W., Schmitt, J.~H.~M.~M., Danner, R., et al.\ 1997, 
\aap, 322, 167

\bibitem[Cassinelli \& Olson(1979)]{CO1979} 
Cassinelli, J.~P., \& Olson, G.~L.\ 1979, \apj, 229, 304

\bibitem[Chu et al.(2001)]{Chu2001} Chu, Y.-H., Guerrero,
  M.~A., Gruendl, R.~A., et al.\ 2001, \apjl, 553, L69

\bibitem[Corradi et al.(1997)]{Corradi1997} Corradi, R.~L.~M.,
  Guerrero, M., Manchado, A., et al.\ 1997, \na, 2, 461

\bibitem[Davis et al.(2012)]{Davis_etal2012}
  Davis, J.~E., Bautz, M.~W., Dewey, D., et al.\ 2012,
  \procspie, 84431A

\bibitem[De Marco et al.(2004)]{dM2004} De Marco, O., Bond, H.~E.,
  Harmer, D., et al.\ 2004, \apjl, 602, L93

\bibitem[Dyson \& Williams(1997)]{Dyson1997} Dyson, J.~E.,
  \& Williams, D.~A.\ 1997, The physics of the interstellar
  medium. Edition: 2nd ed. Publisher: Bristol: Institute of Physics
  Publishing

\bibitem[Fang et al.(2016)]{Fang2016} Fang, X., Guerrero, M.~A.,
  Toal{\'a}, J.~A., et al.\ 2016, \apjl, 822, L19

\bibitem[Fleming et al.(1995)]{Fleming1995} Fleming, T.~A., Schmitt,
  J.~H.~M.~M., \& Giampapa, M.~S.\ 1995, \apj, 450, 401
  
\bibitem[Fruscione et al.(2006)]{Fruscione2006} Fruscione, A.,
  McDowell, J.~C., Allen, G.~E., et al.\ 2006, \procspie, 62701V

\bibitem[Gayley \& Owocki(1995)]{GO1995} 
Gayley, K.~G., \& Owocki, S.~P.\ 1995, \apj, 446, 801

\bibitem[Gruendl et al.(2004)]{Gruendl2004} Gruendl, R.~A., Chu,
  Y.-H., \& Guerrero, M.~A.\ 2004, \apjl, 617, L127

\bibitem[Guerrero(2020)]{Guerrero2020}
Guerrero, M.~A.\ 2020, Proceedings of the Workplans II: Workshop for Planetary Nebula Observations, eds. T.\ Ueta and I.\ Aleman, Galaxies, in prep.\

\bibitem[Guerrero et al.(2019)]{Guerrero2019} 
Guerrero, M.~A., Toal{\'a}, J.~A., \& Chu, Y.-H.\ 2019, \apj, 884, 134

\bibitem[Guerrero et al.(2001)]{Guerrero2001} 
Guerrero, M.~A., Chu, Y.-H., Gruendl, R.~A., et al.\ 2001, \apjl, 553, L55

\bibitem[Guerrero \& De Marco(2013)]{Guerrero2013} Guerrero, M.~A.,
  \& De Marco, O.\ 2013, \aap, 553, A126

\bibitem[Herald \& Bianchi(2011)]{Herald2011} Herald, J.~E., \&
  Bianchi, L.\ 2011, \mnras, 417, 2440

\bibitem[Iping et al.(2002)]{Iping2002} Iping, R.~C., Sonneborn, G.,
  \& Chu, Y.-H.\ 2002, American Astronomical Society Meeting Abstracts
  201, 89.15
  
\bibitem[Kastner et al.(2012)]{Kastner2012} Kastner, J.~H., Montez,
  R., Balick, B., et al.\ 2012, \aj, 144, 58 1997, preprint

\bibitem[Kastner et al.(2000)]{Kastner2000} Kastner, J.~H., Soker, N.,
  Vrtilek, S.~D., et al.\ 2000, \apjl, 545, L57
  
\bibitem[Freeman et al.(2014)]{Freeman2014} Freeman, M., Montez, R.,
  Kastner, J.~H., et al.\ 2014, \apj, 794, 99

\bibitem[Lucy \& White(1980)]{LW1980} 
Lucy, L.~B., \& White, R.~L.\ 1980, \apj, 241, 300

\bibitem[Miszalski et al.(2009)]{Miszalski2009} 
Miszalski, B., Acker, A., Moffat, A.~F.~J., et al.\ 2009, \aap, 496, 813

\bibitem[Montez et al.(2020)]{Montez2020}
  Montez, R., Kastner, J., Jacoby, G., et al.\ 2020,
  American Astronomical Society Meeting Abstracts 52, 307.07

\bibitem[Montez et al.(2010)]{Montez2010}
  Montez, R., De Marco, O., Kastner, J.~H., et al.\ 2010, \apj, 721, 1820

\bibitem[Montez et al.(2015)]{Montez2015} 
  Montez, R., Kastner, J.~H., Balick, B., et al.\ 2015,
  \apj, 800, 8

\bibitem[Montez et al.(2005)]{Montez2005} 
  Montez, R., Kastner, J.~H., De Marco, O., \& Soker, N.\ 2005,
  \apj, 635, 381

\bibitem[Nebot G{\'o}mez-Mor{\'a}n \&
  Oskinova(2018)]{NGO2018} Nebot G{\'o}mez-Mor{\'a}n, A.,
  \& Oskinova, L.~M.\ 2018, \aap, 620, A89

\bibitem[O'Connor et al.(1999)]{OConnor1999} O'Connor, J.~A., Meaburn,
  J., L{\'o}pez, J.~A., et al.\ 1999, \aap, 346, 237

\bibitem[Patriarchi \& Perinotto(1995)]{PP1995} Patriarchi, P., \&
  Perinotto, M.\ 1995, \aaps, 110, 353

\bibitem[Ruiz et al.(2013)]{Ruiz2013} Ruiz, N., Chu, Y.-H., Gruendl,
  R.~A., et al.\ 2013, \apj, 767, 35
  
\bibitem[Sana et al.(2006)]{S2006} 
Sana, H., Rauw, G., Naz{\'e}, Y., et al.\ 2006, \mnras, 372, 661

\bibitem[Steffen et al.(2008)]{Steffen2008} Steffen, M.,
  Sch{\"o}nberner, D., \& Warmuth, A.\ 2008, \aap, 489, 173
  
\bibitem[Toal{\'a} et al.(2019)]{Toala2019} Toal{\'a}, J.~A., Montez,
  R., \& Karovska, M.\ 2019, \apj, 886, 30

\bibitem[Toal{\'a} \& Arthur(2018)]{Toala2018} Toal{\'a}, J.~A., \&
  Arthur, S.~J.\ 2018, \mnras, 478, 1218

\bibitem[Villaver et al.(2003)]{Villaver2003} Villaver, E.,
  Garc{\'\i}a-Segura, G., \& Manchado, A.\ 2003, \apjl, 585, L49
  
\bibitem[Villaver et al.(2012)]{Villaver2012} Villaver, E., Manchado,
  A., \& Garc{\'\i}a-Segura, G.\ 2012, \apj, 748, 94

\bibitem[Yu et al.(2009)]{Yu_etal2009} 
Yu, Y.~S., Nordon, R., Kastner, J.~H., et al.\ 2009, \apj, 690, 440

\bibitem[Zou et al.(2019)]{Zou2019} Zou, Y., Frank, A., Chen, Z., et
  al.\ 2019, arXiv e-prints, arXiv:1912.01647
  
\bibitem[Zucker \& Soker(1993)]{Zucker1993} Zucker, D.~B., \& Soker,
  N.\ 1993, \apj, 408, 579
  
\end{thebibliography}


\end{document}